\documentclass[%
 reprint,
superscriptaddress,
 amsmath,amssymb,
 aps,
 prl,
]{revtex4-2}

\usepackage{graphicx}
\usepackage{float}
\usepackage{graphicx}
\usepackage{dcolumn}
\usepackage{bm}
\usepackage{verbatim}
\usepackage[dvipsnames]{xcolor}
\usepackage{hyperref}
\usepackage[mathlines]{lineno}
\usepackage{makecell}
\usepackage{ulem}
\usepackage{comment}
\begin{document}

\preprint{APS/123-QED}
\title{Quantum light source with lithium tantalate for scalable photonic quantum circuits} 

\author{Yun-Ru Fan}
\thanks{These authors contributed equally to this work}
\affiliation{Institute of Fundamental and Frontier Sciences, University of Electronic Science and Technology of China, Chengdu 611731, China}
\affiliation{Center for Quantum Internet, Tianfu Jiangxi Laboratory, Chengdu 641419, China}
\affiliation{Key Laboratory of Quantum Physics and Photonic Quantum Information, Ministry of Education, University of Electronic Science and Technology of China, Chengdu 611731, China}
\author{Bo-Wen Chen}
\thanks{These authors contributed equally to this work}
\affiliation{State Key Laboratory of Materials for Integrated Circuits, Shanghai Institute of Microsystem and Information Technology, Chinese Academy of Sciences, Shanghai 200050, China}
\affiliation{Center of Materials Science and Optoelectronics Engineering, University of Chinese Academy of Sciences, Beijing 100049, China}
\author{Dan Xu}
\thanks{These authors contributed equally to this work}
\affiliation{Institute of Fundamental and Frontier Sciences, University of Electronic Science and Technology of China, Chengdu 611731, China}
\affiliation{Center for Quantum Internet, Tianfu Jiangxi Laboratory, Chengdu 641419, China}
\affiliation{Key Laboratory of Quantum Physics and Photonic Quantum Information, Ministry of Education, University of Electronic Science and Technology of China, Chengdu 611731, China}
\author{Cheng-Li Wang}
\email{wangcl@mail.sim.ac.cn}
\affiliation{State Key Laboratory of Materials for Integrated Circuits, Shanghai Institute of Microsystem and Information Technology, Chinese Academy of Sciences, Shanghai 200050, China}
\affiliation{Center of Materials Science and Optoelectronics Engineering, University of Chinese Academy of Sciences, Beijing 100049, China}
\author{Hong Zeng}
\affiliation{Institute of Fundamental and Frontier Sciences, University of Electronic Science and Technology of China, Chengdu 611731, China}
\affiliation{Center for Quantum Internet, Tianfu Jiangxi Laboratory, Chengdu 641419, China}
\affiliation{Key Laboratory of Quantum Physics and Photonic Quantum Information, Ministry of Education, University of Electronic Science and Technology of China, Chengdu 611731, China}
\author{Jia-Qi Wang}
\affiliation{Institute of Fundamental and Frontier Sciences, University of Electronic Science and Technology of China, Chengdu 611731, China}
\affiliation{Center for Quantum Internet, Tianfu Jiangxi Laboratory, Chengdu 641419, China}
\affiliation{Key Laboratory of Quantum Physics and Photonic Quantum Information, Ministry of Education, University of Electronic Science and Technology of China, Chengdu 611731, China}
\author{Xu-Qiang Wang}
\affiliation{State Key Laboratory of Materials for Integrated Circuits, Shanghai Institute of Microsystem and Information Technology, Chinese Academy of Sciences, Shanghai 200050, China}
\affiliation{Center of Materials Science and Optoelectronics Engineering, University of Chinese Academy of Sciences, Beijing 100049, China}
\author{Jia-Chen Cai}
\affiliation{State Key Laboratory of Materials for Integrated Circuits, Shanghai Institute of Microsystem and Information Technology, Chinese Academy of Sciences, Shanghai 200050, China}
\affiliation{Center of Materials Science and Optoelectronics Engineering, University of Chinese Academy of Sciences, Beijing 100049, China}
\author{Hai-Zhi Song}
\affiliation{Institute of Fundamental and Frontier Sciences, University of Electronic Science and Technology of China, Chengdu 611731, China}
\affiliation{Southwest Institute of Technical Physics, Chengdu 610041, China}
\author{Hao Li}
\affiliation{State Key Laboratory of Materials for Integrated Circuits, Shanghai Institute of Microsystem and Information Technology, Chinese Academy of Sciences, Shanghai 200050, China}
\author{Li-Xing You}
\affiliation{State Key Laboratory of Materials for Integrated Circuits, Shanghai Institute of Microsystem and Information Technology, Chinese Academy of Sciences, Shanghai 200050, China}
\author{Yan-Yu Wei}
\affiliation{Institute of Fundamental and Frontier Sciences, University of Electronic Science and Technology of China, Chengdu 611731, China}
\author{Kai Guo}
\email{guokai07203@hotmail.com}
\affiliation{Institute of Systems Engineering, AMS Beijing 100141, China}
\author{Xin Ou}
\email{ouxin@mail.sim.ac.cn}
\affiliation{State Key Laboratory of Materials for Integrated Circuits, Shanghai Institute of Microsystem and Information Technology, Chinese Academy of Sciences, Shanghai 200050, China}
\affiliation{Center of Materials Science and Optoelectronics Engineering, University of Chinese Academy of Sciences, Beijing 100049, China}

\author{Guang-Can Guo}
\affiliation{Institute of Fundamental and Frontier Sciences, University of Electronic Science and Technology of China, Chengdu 611731, China}
\affiliation{Center for Quantum Internet, Tianfu Jiangxi Laboratory, Chengdu 641419, China}
\affiliation{Key Laboratory of Quantum Physics and Photonic Quantum Information, Ministry of Education, University of Electronic Science and Technology of China, Chengdu 611731, China}
\affiliation{CAS Center for Excellence in Quantum Information and Quantum Physics, University of Science and Technology of China, Hefei 230026, China}
\author{Qiang Zhou}
\email{zhouqiang@uestc.edu.cn}
\affiliation{Institute of Fundamental and Frontier Sciences, University of Electronic Science and Technology of China, Chengdu 611731, China}
\affiliation{Center for Quantum Internet, Tianfu Jiangxi Laboratory, Chengdu 641419, China}
\affiliation{Key Laboratory of Quantum Physics and Photonic Quantum Information, Ministry of Education, University of Electronic Science and Technology of China, Chengdu 611731, China}
\affiliation{CAS Center for Excellence in Quantum Information and Quantum Physics, University of Science and Technology of China, Hefei 230026, China}

\begin{abstract}
Thin-film lithium tantalate (TFLT) has emerged as a promising integrated photonic platform owing to its low photorefractive noise, high optical damage threshold, and reduced birefringence, attracting increasing interest for scalable photonic technologies. Here, to the best of our knowledge, we demonstrate the first quantum light source with TFLT via spontaneous four-wave mixing, bridging the gap between the rapidly advancing classical TFLT ecosystem and integrated quantum photonics. The fabricated microring exhibits a free spectral range of 350~GHz and an optical quality factor of $10^6$, enabling efficient cavity-enhanced nonlinear interactions. Correlated photon pairs are generated across the telecom band from 1510 to 1570~nm, with a photon pair generation rate of 24 $\mathrm{MHz/mW^{2}}$ at a wavelength of 1535.04 nm. The source delivers strongly antibunched heralded single photons with $g^{(2)}_{H}(0)=0.071\pm0.004$ at a heralding rate of 170 kHz, while the unheralded statistics yield $g^{(2)}(0)=1.93 \pm 0.05$, indicating near-single-temporal-mode emission. Energy-time entanglement is further confirmed by a raw two-photon interference visibility of $92.55\pm0.94\%$, well above the Bell-inequality violation threshold. These results establish TFLT as a manufacturing-compatible platform for scalable photonic quantum circuits, paving the way for the monolithic co-integration of classical and quantum photonic functionalities.

\end{abstract}

\maketitle
\textit{Introduction.}—
The rapid development of thin-film ferroelectric technologies has established a versatile foundation for integrated photonics, with thin-film lithium niobate (TFLN) emerging as a widely adopted platform for high-performance optical signal processing \cite{wang2018integrated, zhang2019broadband, yu2022integrated, nie2025soliton}. Thin-film lithium tantalate (TFLT) has recently attracted growing attention as a complementary member of the ferroelectric family, offering material properties that are highly attractive for scalable photonic technologies \cite{wang2024lithium, niels2026high}. In particular, TFLT exhibits suppressed photorefractive effects, a higher optical damage threshold, and reduced material birefringence, which together facilitate stable operation and simplify the design of polarization-insensitive circuits \cite{wang2025thin, hulyal2025arrayed}. Combined with its compatibility with wafer-scale, high-volume manufacturing processes \cite{wang2024lithium, zhu20258}, these attributes have catalyzed rapid progress in classical TFLT-on-insulator (LTOI) devices. Recent advances include high-speed electro-optic modulators \cite{wang2024ultrabroadband, cai2025heterogeneously}, ultra-high-$Q$ microring resonators (MRRs) with quality factors exceeding $10^7$ \cite{cai2024high, he2025lithium}, and stable octave-spanning dissipative Kerr soliton microcombs \cite{zhu2025octave, cai2025stable}.

Integrated quantum light sources are indispensable building blocks for emerging quantum technologies, including secure quantum communication, large-scale quantum computing, and precision quantum metrology \cite{wang2020integrated, lu2021advances, thomas2024quantum}. Significant progress has been achieved across a wide range of material platforms, including optical fibers \cite{li2005optical}, silica-based microresonators \cite{reimer2016generation, kues2017chip}, silicon nitride \cite{samara2019high, fan2023multi, chen2024ultralow, li2025down}, silicon carbide \cite{rahmouni2024entangled}, lithium niobate \cite{zhao2020high, ma2020ultrabright}, and gallium nitride \cite{zeng2024quantum}. Each platform offers distinct trade-offs in nonlinear efficiency, scalability, and fabrication compatibility, motivating continued exploration of new material systems for integrated quantum photonics. Despite the rapid progress of TFLT in classical photonics, the generation and manipulation of quantum light on this platform remain largely unexplored.

\begin{figure*}[!t]
\centering
\includegraphics[width=18cm]{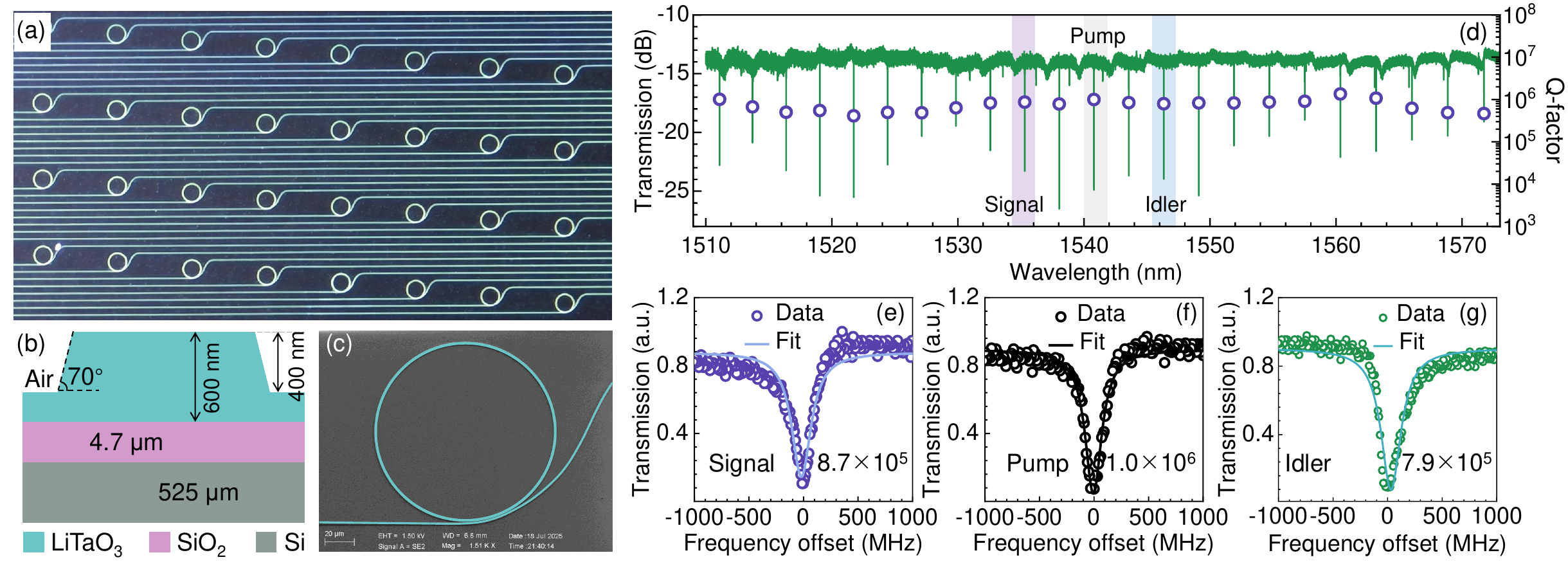}
\caption{Device design and characterization of the LiTaO$_3$ MRR. (a) Optical image of the TFLT chip wtih a series of MRRs, underscoring the potential of the lithium tantalate platform for scalable large-scale quantum photonic circuits. (b) Schematic cross-section of the device, consisting of a 600-nm-thick LiTaO$_3$ layer, a 4.7-$\mu$m-thick SiO$_2$ buffer layer, and a 525-$\mu$m-thick silicon substrate. The waveguide is partially etched by 400 nm with a sidewall angle of about $70^\circ$. (c) Scanning electron microscope (SEM) image of the fabricated MRR. (d) Measured transmission spectrum (green curve) and extracted loaded Q-factors (purple circles) from 1510 to 1570 nm, showing a free spectral range of about 350 GHz. The shaded regions indicate the representative signal, pump, and idler resonances. (e)-(g) Measured resonant curves of the signal, pump, and idler modes, respectively, yielding loaded Q-factors of $8.7\times10^5$, $1.0\times10^6$, and $7.9\times10^5$.}
\label{fig:Fig1}
\end{figure*}

In this work, we report the first demonstration of quantum light generation based on TFLT MRR via the spontaneous four-wave mixing (SFWM) process. By leveraging the high optical quality factor ($Q \approx 10^6$), we achieve the generation of photon pairs across the telecom band (1510--1570~nm) with a photon pair generation rate of 24~$\mathrm{MHz/mW^{2}}$ at a wavelength of 1535.04 nm. The high purity of the source is validated by a measured heralded second-order autocorrelation $g^{(2)}_{H}(0)$ of $0.071 \pm 0.004$, with an unheralded second-order autocorrelation $g^{(2)}(0)$ of $=1.93 \pm 0.05$. Furthermore, we demonstrate energy-time entanglement with a raw two-photon interference visibility of $92.55 \pm 0.94\%$, significantly exceeding the Bell-inequality violation threshold without the need for noise subtraction. These results establish the LTOI platform as a scalable and manufacturing-compatible material system for photonic quantum circuit, opening a path toward monolithic co-integration of classical and quantum photonic functionalities.
\begin{figure*}[t]
\centering
\includegraphics[width=18cm]{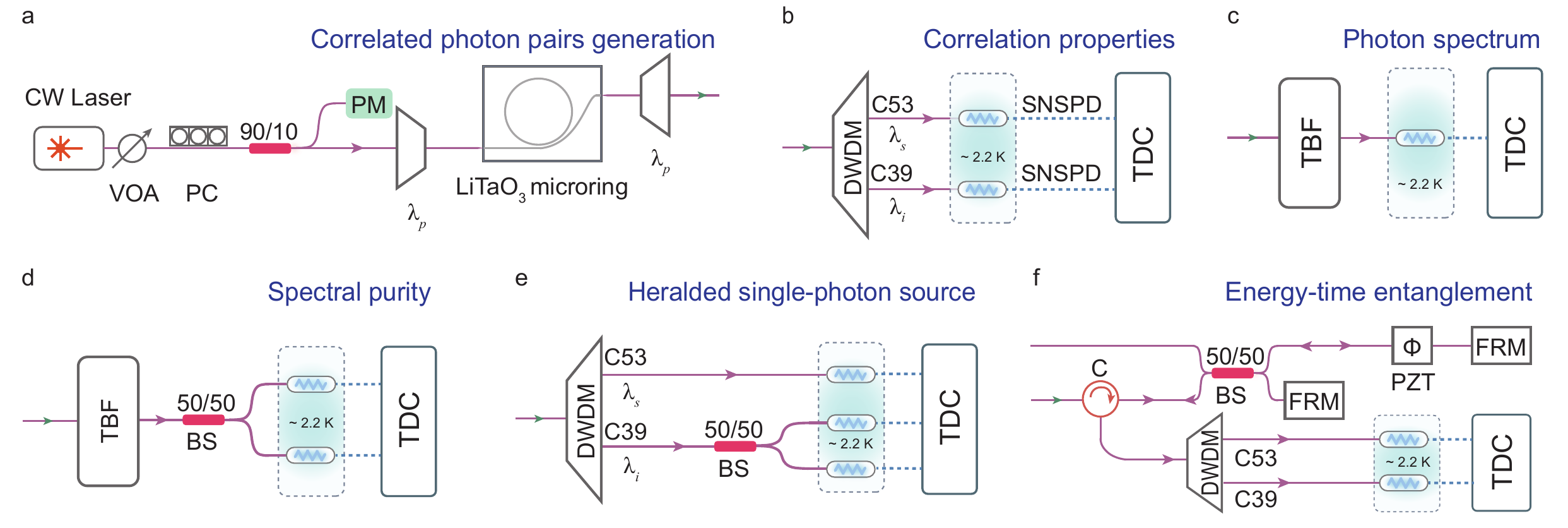}
\caption{Schematic diagram of experimental setups. (a) Generation of correlated photon pairs. (b) Correlation properties. (c) Photon spectrum. (d) Heralded single-photon with HBT experimental setup. (e) Energy-time entanglement with two-photon interference. TL: tunable laser, VOA: variable optical attenuator, PC: polarization controller, BS: beam splitter, PM: power meter, DWDM: dense wavelength division multiplexer, SNSPD: superconducting nanowire single-photon detector, TDC: time-to-digital converter, TBF: tunable bandpass filter, UMI: unbalanced Michelson interferometer. $\lambda_{s}$ and $\lambda_{i}$ are wavelengths of signal and idler photons, respectively. The SNSPDs are operated at a temperature of 2.2 K.}
\label{fig:Fig0}
\end{figure*}

\textit{Device optimization and characterization.}—
Figure~\ref{fig:Fig1}(a) shows the optical image of the TFLT chip wtih a series of MRRs, underscoring the potential of the lithium tantalate platform for scalable large-scale quantum photonic circuits. The device is fabricated on a lithium-tantalate-on-insulator platform consisting of a 600-nm-thick LiTaO$_3$ device layer, a 4.7-$\mu$m-thick SiO$_2$ buffer layer, and a 525-$\mu$m-thick silicon substrate as shown in Fig.~\ref{fig:Fig1}(b). The waveguide is partially etched with an etch depth of 400 nm, leaving a 100-nm slab, and the sidewall angle is measured to be about $70^\circ$. The top widths of the ring waveguide and the bus waveguide are designed to be 1.90 and 1.42 $\mu$m, respectively. The radius of the MRR is 60 $\mu$m, and the gap between the ring and the bus waveguide is 0.45 $\mu$m. Light is coupled into and out of the chip using lensed fibers, with a measured coupling loss of about 7 dB per facet. Figures~\ref{fig:Fig1}(c) show the scanning electron microscope (SEM) images of the fabricated device. Figure~\ref{fig:Fig1}(b) presents the overview of the MRR, while Fig.~\ref{fig:Fig1}(c) shows a magnified image of the coupling region. The pulley-coupling geometry enables efficient coupling between the bus waveguide and the cavity mode. The measured transmission spectrum and the extracted loaded Q-factors from 1510 to 1570 nm are shown in Fig.~\ref{fig:Fig1}(d). The resonances are nearly uniformly distributed over the measured wavelength range, corresponding to a free spectral range (FSR) of about 350 GHz. The loaded Q-factors remain on the order of $10^6$ throughout the telecom band, indicating low propagation loss and favorable cavity enhancement for nonlinear interactions. Transmissions of the signal, pump, and idler modes are shown in Figs.~\ref{fig:Fig1}(e)--\ref{fig:Fig1}(g), respectively, with the loaded Q-factors of $8.7\times10^5$, $1.0\times10^6$, and $7.9\times10^5$, respectively. These results confirm that the fabricated TFLT MRR simultaneously provides high optical quality and appropriate dispersion properties, which are essential for efficient quantum light generation via the SFWM process.


\textit{Photon pair generation and characterization.}—The experimental setup for the photon pair generation and characterization is illustrated in Fig.~{\ref{fig:Fig0}}. As shown in Fig.~{\ref{fig:Fig2}(a), The pump light is provided by a continuous-wave telecom-wavelength laser operating at 1540.29 nm (ITU-C46). Its power and polarization are controlled by a variable optical attenuator (VOA) and a polarization controller (PC), respectively, while a 90:10 beam splitter is used for power monitoring. To suppress the sideband noise of the pump laser and the Raman photons generated in the fiber pigtails, the pump passes through a high-isolation ($\ge 120$ dB) dense wavelength-division multiplexer (DWDM) centered at ITU-C46 before being coupled into the chip through a 10-cm-long lensed fiber pigtail. The light exiting the chip is sent to a pump-rejection filtering module with an isolation of at least 50 dB to remove the residual pump. The generated signal and idler photons at 1535.04 nm (ITU-C53) and 1546.12 nm (ITU-C39), respectively, are then filtered by two DWDMs and detected by superconducting nanowire single-photon detectors (SNSPDs) with a detection efficiency of 90\% and a dark count rate of 50 Hz. The detector outputs are sent to a time-to-digital converter (TDC) for coincidence analysis, as shown in Fig.~\ref{fig:Fig0}(b).

The single side count rate at 1535.04 nm is shown in Fig.~{\ref{fig:Fig2}}(a) at different pump powers. At a pump power of 1.0 mW, the single side count rate of signal photons is 0.78 MHz. The experimental data are represented by circles, while the black line corresponds to the fitted curve using the function $aP_p^2 + bP_p$, where the term $aP_p^2$ accounts for the contribution from correlated photon pairs, and $bP_p$ represents the contribution from noise photons. It can be seen that the quadratic term dominates the photon counting, which suggests the high performance of the generated photon pairs. The coincidence histogram at a pump power of 0.3 mW is shown in Fig.~{\ref{fig:Fig2}(b), with a coincidence window of 2.0 ns - corresponding to the coherence time of photons. Figure~{\ref{fig:Fig2}(c) shows the coincidence, accidental coincidence count rates, and the calculated coincidence-to-accidental ratio (CAR). The photon pair generation rate is calculated to be 24 $\mathrm{MHz/mW^{2}}$ with the collection efficiency of signal and idler photons of 2.3\% and 1.9\%, respectively.  The CAR reaches 158 with a coincidence count rate of 14 Hz, and remains 10 with a coincidence count rate of 11.2 kHz. 

As shown in Fig.~{\ref{fig:Fig0}(c), the single side photon count rates at different resonant modes are measured using a tunable band-pass filter. The measured data are fitted with the function $aP_p^2+bP_p$ to separate the contributions from correlated photons and noise photons, as shown in Fig.~{\ref{fig:Fig2}(d). The fitted contributions are illustrated by the purple and green lines, respectively, indicating correlated photon pairs are successfully generated in a wavelength range from 1510 to 1570 nm. 
\begin{figure*}[t]
\centering
\includegraphics[width=18cm]{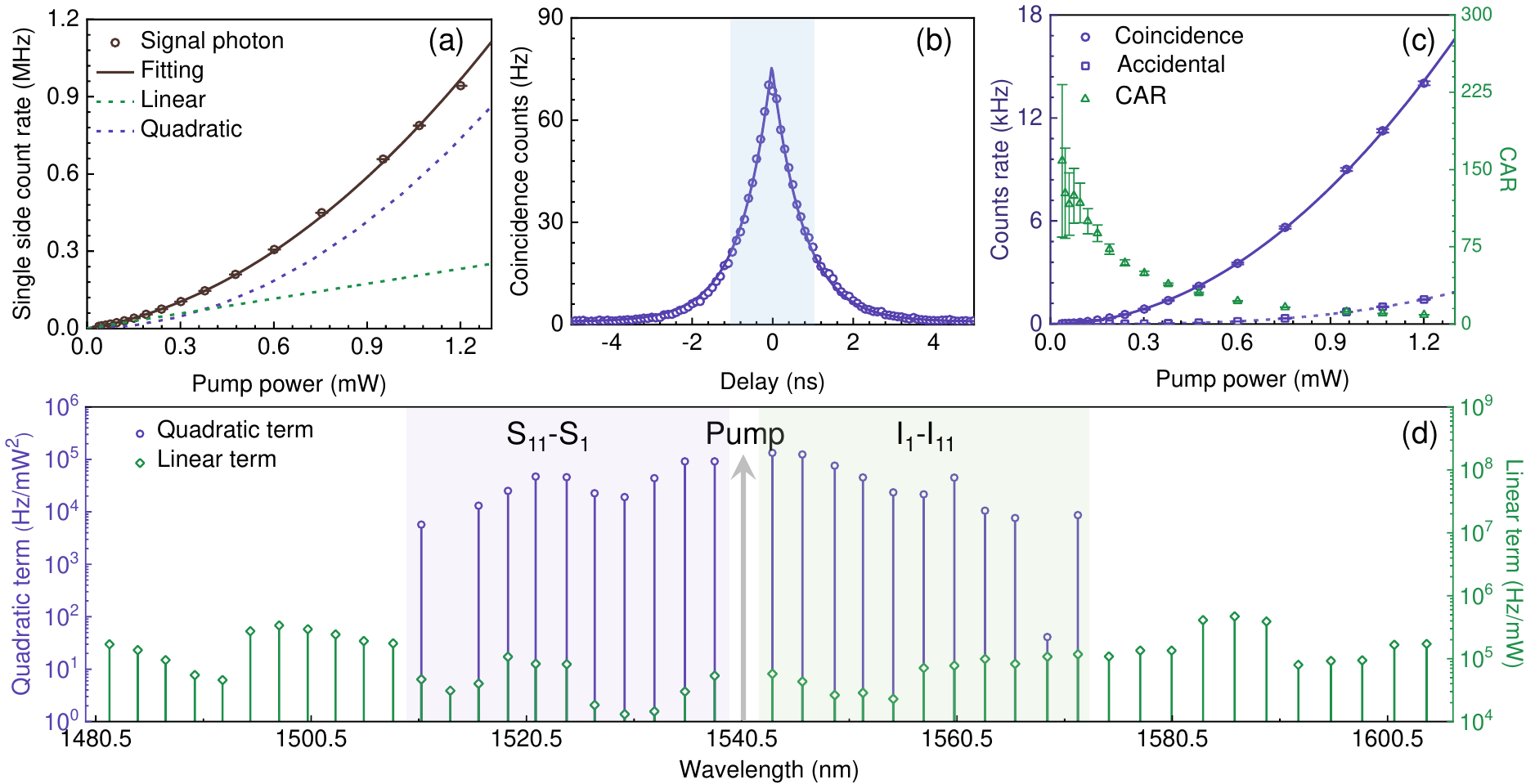}
\caption{Quantum correlations. (a) Single side count rate of signal photons at 1535.04 nm versus pump power. (b) Typical coincidence histogram measured at a pump power of 0.3 mW. (c) Coincidence count rate, accidental coincidence count rate, and calculated CAR. (d) Spectrum of correlated photons and noise photons in the on-resonance case.}
\label{fig:Fig2}
\end{figure*}

\textit{Single-photon purity.}—
The single-photon properties are further characterized, as shown in Fig.~\ref{fig:Fig0}(d). Figure~\ref{fig:Fig3}(a) presents the measured second-order autocorrelation function $g^{(2)}(\tau)$ of the signal photons at 1535.04 nm. By fitting the experimental data, the value of $g^{(2)}(0)$ is extracted to be $1.93 \pm 0.05$, which is close to the ideal value of 2 for a single-mode thermal field. This result indicates that the generated photons exhibit near-single-temporal-mode statistics and thus high single-photon purity. To evaluate the spectral uniformity of the source, the unheralded second-order autocorrelation values of photons at different resonant modes are measured from 1510 to 1570 nm, as shown in Fig.~\ref{fig:Fig3}(b). The measured $g^{(2)}(0)$ values remain close to 2 over the whole wavelength range, confirming that the generated photons preserve the single-mode thermal field property across a broad telecom spectrum.

The heralded single-photon property is further verified by measuring the heralded second-order autocorrelation function $g^{(2)}_{H}(\tau)$ with a Hanbury Brown-Twiss setup. As shown in Fig.~\ref{fig:Fig3}(c), a typical measurement at a detected photon count rate of 170 kHz yields $g^{(2)}_{H}(0)=0.071 \pm 0.004$, showing strong antibunching behavior. This confirms the nonclassical nature and high purity of the heralded single-photon source. Figure~\ref{fig:Fig3}(d) shows the measured $g^{(2)}_{H}(0)$ values under different detected photon count rates. It can be seen that $g^{(2)}_{H}(0)$ gradually increases with the detected photon count rate, which is mainly attributed to the increased multiphoton probability at higher pump powers. Nevertheless, all measured values remain far below 0.5, confirming the high performance of the heralded single-photon source in the lithium tantalate microring resonator.

\begin{figure}[h!]
\centering
\includegraphics[width=8.5cm]{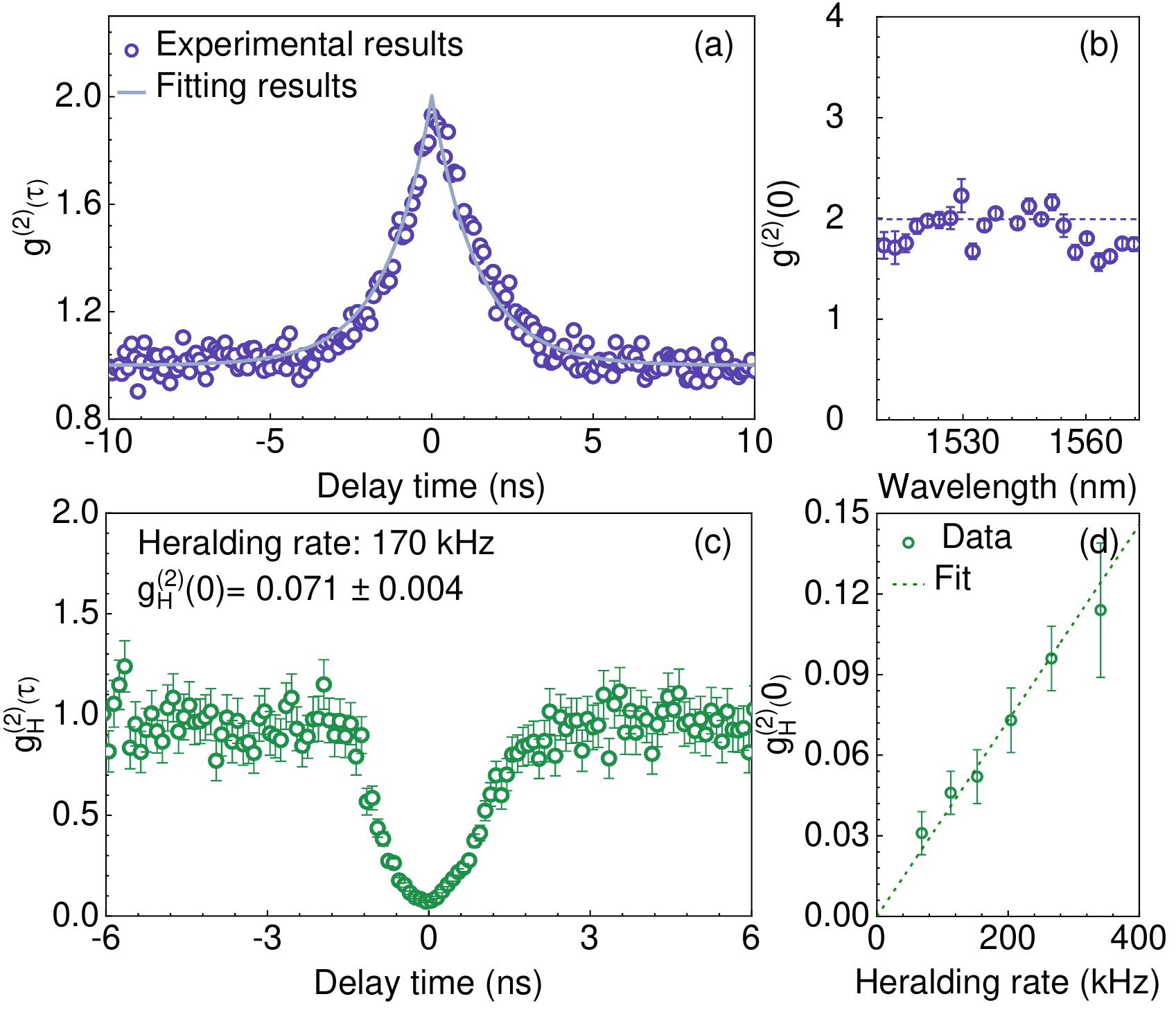}
\caption{Single-photon properties. (a) Measured second-order autocorrelation function $g^{(2)}(\tau)$ of the signal photons at 1535.04 nm. The solid line is the fitting result, giving $g^{(2)}(0)=1.93 \pm 0.05$. (b) Measured $g^{(2)}(0)$ values of photons at different resonant modes from 1510 to 1570 nm. The dashed line indicates the ideal value of 2 for a single-mode thermal field. (c) Measured heralded second-order autocorrelation function $g^{(2)}_{H}(\tau)$ at a heralding rate of 170 kHz, yielding $g^{(2)}_{H}(0)=0.071 \pm 0.004$. (d) Measured $g^{(2)}_{H}(0)$ as a function of the heralding rate.}
\label{fig:Fig3}
\end{figure}

\textit{Energy-time entanglement.}—
The energy-time entanglement of the generated photon pairs is further characterized using an unbalanced Michelson interferometer with a time delay of 10 ns, as shown in Fig.~\ref{fig:Fig0}(e). Figures~\ref{fig:Fig4}(a) and \ref{fig:Fig4}(b) show the coincidence histograms measured under constructive and destructive interference conditions, respectively. Under the constructive condition, the central coincidence peak is significantly enhanced, while under the destructive condition it is strongly suppressed. The side peaks remain nearly unchanged in both cases, indicating that only the indistinguishable two-photon amplitudes contribute to the interference of the central peak. Figure~\ref{fig:Fig4}(c) shows the single-photon interference by scanning the phase of the interferometer. Figure~\ref{fig:Fig4}(d) shows the Franson interference fringe obtained from the central coincidence counts as a function of the phase. The coincidence counts exhibit a clear sinusoidal variation, with a period equal to half that of the single-photon interference. By fitting the measured coincidence counts, a raw two-photon interference visibility of $92.55 \pm 0.94\%$ is obtained with 1000-time Monte Carlo methods. This visibility is well above the Bell-inequality violation threshold, confirming the high-quality energy-time entanglement of the generated photon pairs. The single side count rates of the signal and idler photons, shown on the right axis, remain nearly constant during the phase scan. 
\begin{figure}[t]
\centering
\includegraphics[width=8.5cm]{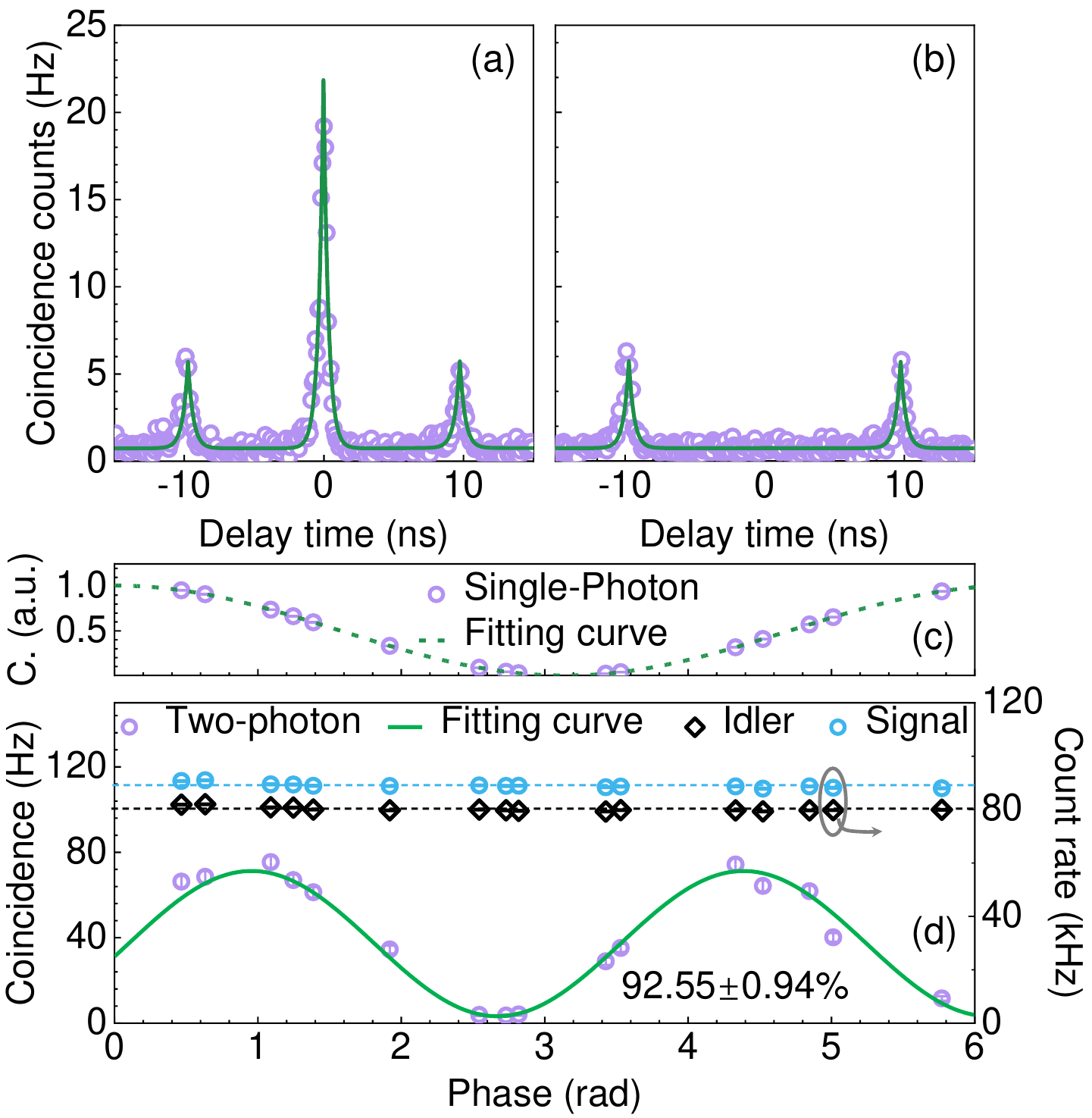}
\caption{Energy-time entanglement property. (a) Coincidence histogram measured under constructive interference. (b) Coincidence histogram measured under destructive interference. (c) Measured single-photon interference fringe as a function of the phase. (d) Franson interference fringe of the central coincidence counts as a function of the phase. The solid curve is the sinusoidal fitting result, yielding a raw two-photon interference visibility of $92.55 \pm 0.94\%$. The right axis shows the single side count rates of the signal and idler photons during the phase scan, which remain nearly unchanged.}
\label{fig:Fig4}
\end{figure}

\textit{Discussion and conclusion.}—
In summary, we have demonstrated, to the best of our knowledge, the first quantum light source with a TFLT MRR via the SFWM process. The fabricated device exhibits loaded Q-factors on the order of $10^6$, enabling high photon-pair generation of 24 $\mathrm{MHz/mW^{2}}$. Broadband photon-pair generation is observed from 1510 to 1570 nm, indicating that the TFLT microring provides suitable cavity enhancement and modal properties for integrated quantum light generation across the telecom band. In addition, the generated photons exhibit near-single-temporal-mode statistics with $g^{(2)}(0)$ approaching the single-mode thermal limit, while the heralded photons show strong antibunching with $g^{(2)}_{H}(0)=0.071 \pm 0.004$. High-quality energy--time entanglement is further confirmed by a raw Franson interference visibility of $92.55 \pm 0.94\%$ without background subtraction.

Our work highlights the unique potential of the lithium-tantalate-on-insulator platform for integrated quantum photonics. Compared with conventional TFLN, TFLT offers weaker photorefractive effects, a higher optical damage threshold, and reduced birefringence, which are favorable for stable cavity operation, broadband nonlinear interactions, and scalable photonic integration. These material advantages, together with its compatibility with wafer-scale and high-volume manufacturing, make TFLT particularly attractive for the development of manufacturable quantum photonic chips. In this sense, the present work extends the rapidly developing classical TFLT ecosystem into the quantum regime and establishes a missing building block for a comprehensive ferroelectric photonic platform.

There remains substantial room for further performance improvement. The detected photon rates can be significantly increased by reducing the chip-to-fiber coupling loss and the off-chip filtering loss. The CAR can be further improved by suppressing parasitic noise, including Raman photons induced in the fiber components and excess scattering associated with fabrication imperfections. Further optimization of material quality, device geometry, and cavity coupling conditions should also enable higher brightness, lower noise, and broader spectral coverage. Combined with the already established capabilities of TFLT in high-speed electro-optic modulation, microcomb generation, and scalable PIC fabrication, the demonstrated quantum light source opens a path toward monolithic co-integration of classical and quantum photonic functionalities on a single chip, providing a promising route for manufacturing-compatible quantum photonic circuits.

\begin{acknowledgments}

\textbf{Acknowledgments}

This work was supported by Quantum Science and Technology-National Science and Technology Major Project (Nos.~2024ZD0300800, 2021ZD0301702), National Natural Science Foundation of China (Nos.62293521, 62475039, 62405046), Sichuan Science and Technology Program (Nos.~2024YFHZ0370, 2024YFHZ0369, 2024YFHZ0368, 2026NSFSC1439), Shanghai Science and Technology Program (No.~24CL2901000), Strategic Priority Research Program of Chinese Academy of Sciences (No.~XDB1440200), Tianfu Jiangxi Laboratory (No.~TFJX-ZD-2025-005).


\textbf{Data availability}

All data needed to evaluate the conclusions in the paper are present in the paper. Additional data related to this paper may be requested from the authors.

\textbf{Conflict of interest} 

The authors declare no competing interests.

\textbf{Author contributions} 

Y.-R.F., B.-W.C., and D.X. mainly carried out the experiment. H.Z., J.-Q.W., H.-Z.S., and Y.-Y.W. assisted in the acquisition of experimental data. Y.-R.F., D.X., and Q.Z. analyzed the data and wrote the manuscript. B.-W.C., C.-L.W., X.O., X.-Q.W, and J.-C.C. fabricated the device. H.L. and L.-X.Y. developed and maintained the SNSPDs used in the experiment. C.-L.W., K.G., X.O., G.-C.G., and Q.Z. conceived and supervised the project. 

\end{acknowledgments}

%

\end{document}